\documentclass[twocolumn,  amsmath, amssymb, aps, prb,  superscriptaddress,
]{revtex4-1}%
\usepackage{graphicx}
\usepackage{dcolumn}
\usepackage{bm}
\usepackage{amsmath}
\usepackage{amsfonts}
\usepackage{amssymb}
\usepackage{natbib}%
\providecommand{\U}[1]{\protect\rule{.1in}{.1in}}
\begin{document}

\begin{abstract}

\end{abstract}

%

\preprint{Revised version for 3rd review}%
%

\title
{Josephson coupling through one-dimensional ballistic channel in semiconductor-superconductor hybrid quantum point contacts}%
%

\author{Hiroshi Irie}%
%

\email{irie.hiroshi@lab.ntt.co.jp}%
%

\author{Yuichi Harada}%
%

\affiliation{
NTT Basic Research Laboratories, NTT Corporation, 3-1 Morinosato-Wakamiya, Atsugi 243-0198, Japan}%
%

\author{Hiroki Sugiyama}%
%

\affiliation{
NTT Photonics Laboratories, NTT Corporation, 3-1 Morinosato-Wakamiya, Atsugi 243-0198, Japan}%
%

\author{Tatsushi Akazaki}%
%

\affiliation{
NTT Basic Research Laboratories, NTT Corporation, 3-1 Morinosato-Wakamiya, Atsugi 243-0198, Japan}%
%

\noaffiliation
%

\date{\today}%
%

\begin{abstract}%

We study a superconducting quantum point contact made of a narrow In$_{0.75}%
$Ga$_{0.25}$As channel with Nb proximity electrodes. The narrow channel is
formed in a gate-fitted constriction of InGaAs/InAlAs/InP heterostructure
hosting a two-dimensional electron gas. When the channel opening is varied
with the gate, the Josephson critical current exhibits a discretized variation
that arises from the quantization of the transverse momentum in the channel.
The quantization of Josephson critical current persists down to the
single-channel regime, providing an unambiguous demonstration of a
semiconductor--superconductor hybrid Josephson junction involving only a
single ballistic channel.%

\begin{description}%
%

\item[PACS numbers]%
74.45.+c, 03.75.Lm, 73.63.Nm, 71.70.Ej%

\end{description}%
%

\end{abstract}%
%

\maketitle

\section{\label{sec:level1}Introduction}

The conductance of a ballistic point contact linking two reservoirs in thermal
equilibrium is quantized in multiples of the conductance quantum $G_{0}%
=2e^{2}/h$\cite{ref01, ref02}. The origin of this phenomenon is the
quantization of the transverse momentum in the narrow constriction.
Strikingly, $G_{0}$ is independent of the parameters characterizing the
contact, and the conductance is thus solely determined by the number of modes.
Another interesting feature is that the contact has a finite resistance even
though no scattering is assumed at the constriction. What happens if we
replace the reservoirs with superconductors? This question was answered by
Beenakker and van Houten a few years after the discovery of the conductance
quantization\cite{ref03}. They theoretically analyzed a superconducting
quantum point contact (SQPC) made of a smooth and impurity-free
superconducting constriction, and they showed that superconducting Josephson
current is carried through the Andreev bound states which are phase-coherent
discrete levels formed in each quantized mode. Since the energy spectrum of
these levels is insensitive to the junction properties in the short-channel
limit, the Josephson critical current $I_{\text{c}}$ per mode is described by
the junction-independent parameters as $I_{0}=e\Delta_{0}/\hbar$ ($\Delta_{0}$
is the superconducting gap). Note that $I_{0}$ has $\Delta_{0}$ in its form in
addition to the fundamental physical constants. In this respect, in contrast
to the conductance quantization, the phenomenon is not universal. More
theoretical work has been undertaken to take into account more realistic
situations, such as a constriction longer than the superconducting coherence
length, a Schottky barrier at the interface between the constriction and
reservoirs, and elastic scattering in the constriction\cite{ref04a, *ref04b,
ref05, ref06}. These analyses clarified that the step-like variation of
$I_{\text{c}}$ as a function of the mode number survives in a wide range of
junction parameters, despite the fact that $I_{\text{c}}$ per mode is
sensitively altered by the junction geometry and scattering process.

To prove the quantization of $I_{\text{c}}$, two types of experiments have
been undertaken. The first uses a mechanically controllable break junction
(MCBJ) made of two superconducting banks bridged by an atomically narrow
constriction. By mechanically elongating or contracting the structure, the
constriction's diameter, and hence the number of transport modes, can be
tuned. A discretized change of the superconducting critical current with a
step size comparable to $e\Delta_{0}/\hbar$ was observed in a Nb
MCBJ\cite{ref07}. Further study using different superconducting materials
revealed that atomic valence orbitals constitute the current-carrying
channels\cite{ref08}. This finding indicates that it is difficult to
manipulate either the number of transport modes or their transmission
probabilities in a controlled way because these channels are extremely
sensitive to the atomic configuration. Moreover, considering that even a
single atom has several valence orbitals, isolating a single conducting
channel is a challenging task for most metals except monovalent metals like
Au\cite{Scheer}.

The second approach exploits the semiconductor--superconductor (Sm--Sc) hybrid
structure. In a Sc/Sm/Sc junction, Josephson coupling is attained via Andreev
reflection (AR) of quasiparticles confined in the Sm region. The transport
properties of the quasiparticles, and hence the Josephson-junction (JJ)
characteristics, can be controlled by means of the external electric field
from a gate electrode. Quantum point contact (QPC) in a high electron mobility
two-dimensional electron gas (2DEG)\cite{ref04a, *ref04b, ref09} and
gate-fitted nanowires\cite{Xiang, Doh, Nishio} have been used as Sm materials
to induce quantized conducting channels. In the former case, ballistic
one-dimensional (1D) channels with almost perfect transmission can be formed
with comparative ease because of their long mean free path. The number of
channels is electrically tunable by a gate, which offers better
controllability than an MCBJ. Takayanagi \textit{et al.}\cite{ref09}
experimentally demonstrated a 2DEG-based hybrid SQPC that utilizes an
InAs-based QPC with Nb proximity electrodes. They showed a stepwise change of
both $I_{\text{c}}$ and normal-state conductance $G_{\text{n}}$ as a function
of gate voltage. However, the quantization steps were vaguely visible, and the
operation was limited to a few-channel regime ($n>4$, where $n$ is the number
of 1D channels). Later, in a follow-up study\cite{ref10}, it was found that
$I_{\text{c}}$ is excessively suppressed for $n<3$, which hinders access to
the single-channel operation. Other than the two reports above, there are no
previous reports on this subject, and the realization of single-channel
operation has been unattained so far. As for the nanowire-based hybrid SQPC,
the quantized steps of $I_{\text{c}}$ and $G_{\text{n}}$ have not been
observed in most devices using InAs nanowires\cite{Doh, Nishio} except for the
one using a Ge/Si core/shell nanowire with Al electrodes\cite{Xiang}. In the
present paper, we present a 2DEG-based hybrid SQPC that exhibits staircase
variation of $I_{\text{c}}$ from the multiple-channel regime down to the
single-channel regime, which provides compelling evidence of the quantization
of the Josephson critical current.

\section{Experiment}

\begin{figure}
[ptb]
\begin{center}
\includegraphics[
width=2.655in
]%
{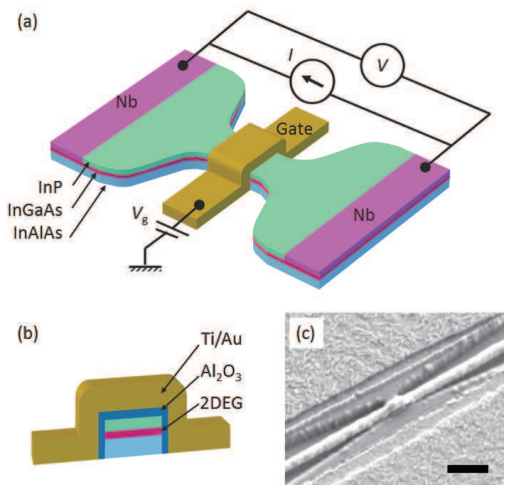}%
\caption{(Color online) Schematic drawings of the (a) SQPC and (b) cross
section of the wrap-gate QPC. (c) Scanning electron micrograph of a
representative device tested. The black scale bar is 300 nm.}%
\label{fig01}%
\end{center}
\end{figure}

The device structure of the SQPC studied is illustrated in Fig. \ref{fig01}.
Similar to the SQPCs in the previous studies\cite{ref09, ref10}, it consists
of a 2DEG-based QPC formed in a high-In-content InGaAs and two Nb electrodes
in the vicinity of the QPC. In what follows, we describe the device structure,
focusing on two major modifications to the SQPC studied previously.%

Instead of the conventional finger-gate geometry, our QPC has a wrap-gate
geometry, in which a narrow constriction made of an InP/In$_{0.75}$Ga$_{0.25}%
$As/InAlAs inverted-type high-electron-mobility transistor (HEMT)
structure\cite{ref11} is wrapped with an Al$_{2}$O$_{3}$ insulator and Ti/Au
gate. The Al$_{2}$O$_{3}$ layer was formed by the atomic layer deposition
technique, which provides low interface state density, resulting in a good
gate controllability. Our previous study demonstrated a well-behaved QPC
operation\cite{ref12}, i.e., conductance steps with a constant stepheight of
$G_{0}$. More importantly, such clear conductance quantization is sustained
even at temperatures down to 0.3 K and at zero magnetic field, where quantized
steps are easily distorted due to scattering around the QPC\cite{Schapers1,
Ramvall}. The geometry of the QPC used here is the same as the one in Ref.
\citenum{ref12}%
. The width and length of the narrow constriction are 120 and 200 nm,
respectively. The length of the QPC is 80 nm, which is defined by the width of
the gate electrode. Electron mobility $\mu_{\text{e}}$ and density
$n_{\text{s}}$ of the 2DEG at 1.9 K are 156,000 cm$^{2}$/Vs and 1.9$\times
$10$^{12}$ 1/cm$^{2}$, respectively. The calculated elastic mean free path
$l_{\text{e}}$ ($=$ $\hbar\mu_{\text{e}}/e\sqrt{2\pi n_{\text{s}}}$) is 3.5 $%
\operatorname{\mu m}%
$. Electron effective mass $m^{\ast}$ is obtained from the temperature
dependence of the Shubnikov-de Haas oscillation as $0.043m_{\text{0}}$, where
$m_{\text{0}}$ is the electron rest mass. Further information regarding to the
wrap-gate QPC can be found in Ref.
\citenum{ref12}%
.

The Nb electrode, another key component of the SQPC, is fabricated by a
lift-off process employing electron beam lithography. In order to achieve high
AR probability, the Nb electrode has to directly touch the 2DEG, and the
formation of a potential barrier at the interface should be avoided. A
combination of coarse wet etching and subsequent \textit{in situ} Ar plasma
cleaning is carried out before Nb deposition. The former uses a phosphoric
acid solution to selectively etch the InP layer on the InGaAs 2DEG layer.
Thanks to the self-terminating process, the time required for the following
\textit{in situ} plasma cleaning can be minimized. The distance between the
two Nb electrodes $L_{\text{ch}}$ is chosen to be 300 nm. Since $L_{\text{ch}%
}$ is much shorter than $l_{\text{e}}$, the SQPC is in the ballistic regime.
When we discuss the superconducting properties of the SQPC, $L_{\text{ch}}$
should also be compared with the coherence length in the 2DEG $\xi_{0}$
($=\hbar v_{\text{f}}/\pi\Delta_{0}$, where $v_{\text{f}}$ and $\Delta_{0}$
are the Fermi velocity and the superconducting gap). If $\Delta_{\text{Nb}%
}=1.27$ meV is used for $\Delta_{0}$, which is calculated from the critical
temperature of the Nb electrode ($T_{\text{c}}=8.4$ K), we obtain $\xi
_{0}=164$ nm, giving $L_{\text{ch}}/\xi_{0}>1$. This indicates that the SQPC
is categorized as a long junction, in which multiple Andreev levels lie in the
SQPC for each transport mode\cite{ref13}. However, as we will discuss in the
next section, $\Delta_{0}$ will be replaced with a smaller value when a
minigap is induced in the 2DEG via the superconducting proximity effect. In
this case, the SQPC is in the short-channel regime ($L_{\text{ch}}/\xi_{0}<1$).

Current-voltage (\textit{I-V}) characteristics of the SQPC were measured in a
dilution refrigerator. Unless otherwise stated, all measurements were
performed at 20 mK. A current source with a 200 k%
$\Omega$
load resistor was used for the bias sweep. The bias voltage $V$ was measured
by the four-terminal mehod using two independent contacts for each Nb
electrode, which eliminated parasitic voltage drops other than that of SQPC.
All the electrical lines inside the dilution refrigerator were twisted pairs
equipped with a two-stage filter installed at the mixing chamber. The first
stage comprises resistance-capacitance filters with a cutoff frequency of
approximately 20 kHz. The second one consists of 2-m-long twisted pairs of
constantan lines sealed tightly within folded copper tape. The latter filters
out the high-frequency noise (frequency above 1 GHz), which is crucial for
correctly evaluating the \textit{I-V} curve of a JJ\cite{ref14, ref15}. In
order to analyze the effects of the mode quantization, \textit{I-V} curves
were recorded at different values of gate voltage $V_{\text{g}}$. For each
$V_{\text{g}}$, 20 measurements were performed to average out the statistical variation.

\bigskip

\section{Josephson junction characteristics of SQPC}

\begin{figure}
[ptb]
\begin{center}
\includegraphics[
width=3.2785in
]%
{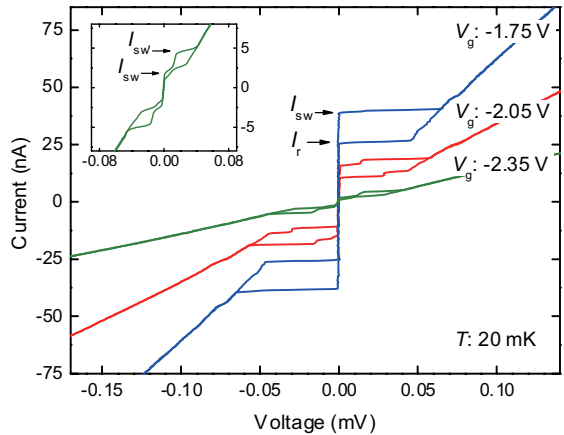}%
\caption{(Color online) \textit{I-V} characteristics of the SQPC taken at
three representative values of $V_{\text{g}}$. The inset shows a magnified
view of the \textit{I-V} curve with $V_{\text{g}}=-2.35$ V.}%
\label{fig02}%
\end{center}
\end{figure}

Figure \ref{fig02} shows typical \textit{I-V} characteristics for three
representative $V_{\text{g}}$ values. All \textit{I-V} curves show JJ
characteristics with a superconducting branch at zero voltage. We define the
current at which a finite voltage appears in a forward sweep as switching
current $I_{\text{sw}}$. In a reverse sweep, the finite-voltage (resistive)
state goes back to the superconducting branch at retrapping current
$I_{\text{r}}$. The obtained \textit{I-V} curves show a hysteresis;
$I_{\text{sw}}$ is not equal to $I_{\text{r}}$, which is commonly seen in
Sc/Sm or Sc/metal hybrid Josephson junctions at temperatures much below the
superconducting critical temperature. The hysteresis is most likely caused by
heating in the resistive state. A direct measurement of normal-metal
electronic temperature in an Al/Cu JJ demonstrates that the temperature
increases once the JJ switches from a superconducting state to a resistive
state which results in the reduction of $I_{\text{r}}$\cite{Courtois}.
Moreover, Sc/Sm hybrid JJs in general show a temperature independent
$I_{\text{r}}$ in the temperature range where its \textit{I-V} curve shows
hysteresis\cite{Krasnov, Deon, Schapers2}, indicating that $I_{\text{r}}$ is
suppressed in low-temperature limit due to the increased temperature by
heating. As shown later in Fig. \ref{fig04}, our SQPC also shows a temperature
independent $I_{\text{r}}$, supporting the idea that heating is the origin of
the hysteresis in our SQPC. We note, however, that some controversial results
have been reported regarding the origin of the hysteresis. A comparison
between Sc/Sm/Sc junctions with and without a shunt capacitance indicates that
the hysteresis is predominantly due to the underdamped nature of the
JJ\cite{Krasnov}. Although the intrinsic quality factor of the junction itself
is very small because of the small capacitive coupling between in-plane Sc
electrodes, a stray capacitance could enhance the quality factor considerably,
causing the underdamped behavior\cite{ref14}. For the underdamped JJ, we
should be aware that a small fluctuation drives the JJ to switch to the
resistive state below the intrinsic critical current, leading to a measured
$I_{\text{sw}}$ lower than the theoretical $I_{\text{c}}$. Nevertheless, we
hereafter assume that our JJ is in the overdamped regime and that the
hysteresis is caused by heating.%

With regard to the $V_{\text{g}}$ dependence of the \textit{I-V} curves in
Fig. \ref{fig02}, both $I_{\text{sw}}$ and $I_{\text{r}}$ decrease with
decreasing $V_{\text{g}}$, and the slope of the \textit{I-V} curves in the
resistive state simultaneously changes with $V_{\text{g}}$. This gate
controllability in terms of superconducting properties is a unique feature of
the Sc/Sm hybrid JJ. We also notice that the \textit{I-V} curves for
$V_{\text{g}}$ $=-2.05$ and $-2.35$ V have an additional shoulder at the
superconducting-to-resistive transition. A magnified view of the \textit{I-V}
curve for $V_{\text{g}}=-2.35$ V shown in the inset of Fig. \ref{fig02}
clearly display the shoulder at approximately 15 and 30 $%
\operatorname{\mu V}%
$ for the forward and reverse sweep respectively. At elevated temperatures and
under some $V_{\text{g}}$ value, more than two shoulders appear in a single
bias sweep (data not shown). These shoulders are caused by the ac Josephson
effect \cite{ref16, ref29} due to an unintentionally formed cavity in the
measurement system. As a result, the shoulders appear at integer multiples of
15 $%
\operatorname{\mu V}%
$ (corresponding to resonant frequency of 7.3 GHz), and these positions are
the same for all three samples tested. Although this effect seems to suppress
the switching current of the first transition, we take it as $I_{\text{sw}}$
as depicted in the inset of Fig. \ref{fig02}. Since this suppression occurs
for $I_{\text{sw}}$ below approximately $20$ nA at 20 mK, it does not affect
the \textit{I-V} characteristics for most $V_{\text{g}}$ values except near
the pinch-off.

\begin{figure}
[ptb]
\begin{center}
\includegraphics[
width=3.1401in
]%
{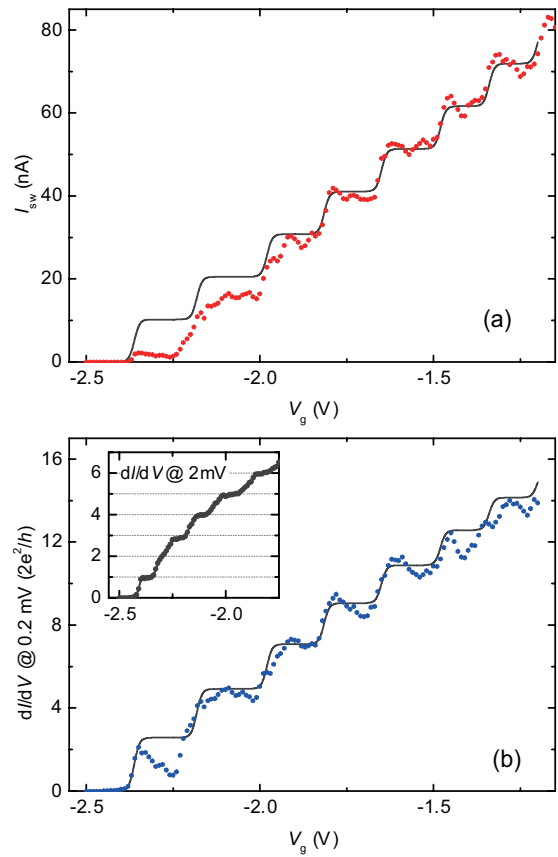}%
\caption{(Color online) (a) $V_{\text{g}}$ dependence of $I_{\text{sw}}$. (b)
$V_{\text{g}}$ dependence of d$I$/d$V$ taken at $V=0.2$ mV. For both (a) and
(b), the dots represent experimental data, while the lines are fitted curves.
Refer to the main text for the details of the fitting. The inset in (b) shows
d$I$/d$V$ taken at $V=2$ mV. Note that the data in the inset were taken in a
cool down cycle different from that in the main panels.}%
\label{fig03}%
\end{center}
\end{figure}

\begin{figure}
[ptb]
\begin{center}
\includegraphics[
width=3.1583in
]%
{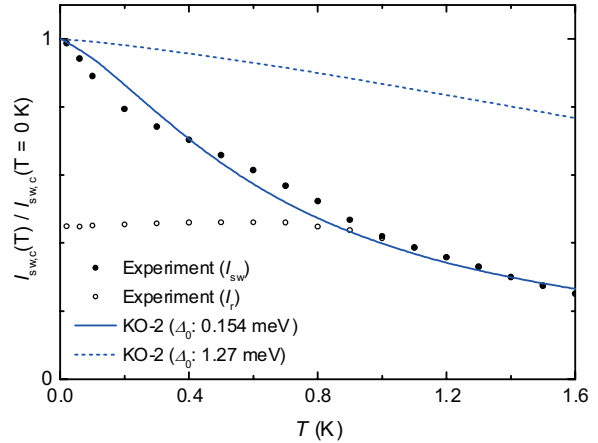}%
\caption{(Color online) Temperature dependence of the switching current
normalized by the value at 0 K. The solid and open circles are experimental
data of $I_{sw}$ and $I_{r}$, respectively. The solid and dashed lines are the
curves calculated using the KO-2 model with $\Delta_{0}=0.154$ meV and
$\Delta_{0}=1.27$ meV, respectively.}%
\label{fig04}%
\end{center}
\end{figure}

To study the effects of transport mode quantization, we show the $V_{\text{g}%
}$ dependence of $I_{\text{sw}}$ and differential conductance d$I/$d$V$ in
Figs. \ref{fig03}(a) and \ref{fig03}(b), respectively. The d$I/$d$V$ is
obtained by numerically differentiating an \textit{I-V} curve at bias voltage
$V=0.2$ mV. The solid lines represent lines fitted using the following equations:%

\begin{align}
&  I_{\text{sw}}=I_{\text{sw}0}\underset{n=1}{\sum}T_{n}(V_{\text{g}%
}),\nonumber\\
&  \left.  \frac{\text{d}I}{\text{d}V}\right\vert _{V}^{-1}=\left[
g_{0}\underset{n=1}{\sum}T_{n}(V_{\text{g}})\right]  ^{-1}+R_{\text{c}%
},\label{eq01}%
\end{align}
\linebreak where $I_{\text{sw0}}$ is the switching current per channel,
$g_{0}$ is the conductance per channel, $R_{\text{c}}$ is the contact
resistance at the Nb/InGaAs interface, and $T_{n}(V_{\text{g}})$ is the
transmission probability of the $n$th channel. We presume here that each 1D
channel contributes to the total switching current (or the conductance) by an
equal amount of $I_{\text{sw0}}$ (or $g_{0}$) and the switching current (or
the conductance) per channel is linearly dependent on the transmission
probability. A saddle-point model is used to describe the potential landscape
at QPC\cite{ref17}, providing $T_{n}(V_{\text{g}})=\left\{  1+\exp\left[
-2\pi(E_{\text{f}}(V_{\text{g}})-E_{n})\right]  /\hbar\omega_{x}\right\}
^{-1}$, where $E_{n}$ is the lowest energy of the $n$-th subband, $\hbar
\omega_{x}$ is the curvature of the saddle-point potential parallel to the
current flow, and $E_{\text{f}}(V_{\text{g}})$ is the $V_{\text{g}}$-dependent
Fermi level. For $E_{\text{f}}(V_{\text{g}})$, we use the relation with a
constant d$E/$d$V_{\text{g}}$ ($=0.127$ eV/V)\cite{ref12}. To calculate the
fitted lines shown in the Figs. \ref{fig03}(a) and \ref{fig03}(b), we used the
following set of numbers: ($I_{\text{sw0}}$, $g_{0}$, $R_{\text{c}}$,
$\hbar\omega_{x}$) = (10.3 nA, 2.7 $G_{0}$, 230 $\Omega$, 5.9 meV).%

The main finding of the present work is the clear stepwise variation of
$I_{\text{sw}}$ with respect to $V_{\text{g}}$ in Fig. \ref{fig03}(a), by
which we unambiguously prove the Josephson coupling through the quantized 1D
channels. Furthermore, the reasonable fit with $n$-independent step height
$I_{\text{sw0}}=10.3$ nA (except for $n=1$ and $2$) demonstrates both an equal
contribution of each 1D channel to $I_{\text{sw}}$ and negligible intermixing
between the channels, which is consistent with existing theories. The
suppressed stepheights for $n=1$ and 2 are presumably caused by the thermal
activation escape because the Josephson coupling energy is comparable to the
thermal energy of the bath temperature. The thermal activation escape in
overdamped JJs causes the so-called phase diffusion\cite{Mel'nikov} and
results in a rounded switching in the \textit{I-V} curve, which is indeed seen
in the inset of Fig. \ref{fig02}. Despite the suppression of $I_{\text{sw}}$
for the first channel, JJ behavior with an accompanying superconducting branch
is clearly observed when $V_{\text{g}}$ is set such that the SQPC holds a
single ballistic channel (see the inset of Fig. \ref{fig02} for the
\textit{I-V} curve).

In what follows, experimental $I_{\text{sw0}}$ is compared with theoretical
$I_{\text{c}}$. In the simplest model assuming an SQPC in the short-channel
limit ($L_{\text{ch}}/\xi_{0}<<1)$ with an ideal Sc/Sm interface,
$I_{\text{c}}$ is equal to $e\Delta_{0}/\hbar$\cite{ref03, ref04a, *ref04b}.
Given the influence of a finite channel length and a Schottky barrier at the
Sc/Sm interface, $I_{\text{c}}$ is modified as\cite{ref05}%

\begin{equation}
I_{c}=\alpha\frac{e}{\tau}, \label{eq02}%
\end{equation}
with%

\[
\tau=\frac{\hbar}{\Delta_{0}}+\tau_{0}\left(  \frac{2}{D}-1\right)
\]
where $\alpha$ is a coefficient determined from the Fabry-P\'{e}rot-type
interference effect due to the normal reflection at the Sc/Sm
interface\cite{ref05}, $\tau_{0}$ is the time of flight of a quasiparticle in
2DEG $L_{\text{ch}}/v_{\text{f}}$, and $D$ is the tunneling probability at the
Sc/Sm interface. Note that Eq. (\ref{eq02}) is reduced to the simplest form
when $\alpha=1$, $\tau_{0}\rightarrow0$ and $D\rightarrow1$ are assumed.
According to Ref.
\citenum{ref03}%
, the value of $\alpha$ oscillates as a function of $V_{\text{g}}$, as a
result of the interference condition, within a range between 1 (constructive
interference) and $D/4\pi$ (destructive interference). The value of $D$ is
determined from $D=1/(1+Z^{2})$, where $Z$ represents dimensionless barrier
strength. $Z$ can be roughly estimated from the relation $R_{\text{N}%
}=R_{\text{Sh}}(1+2Z^{2})$, where $R_{\text{N}}$ and $R_{\text{Sh}}$ are the
normal resistance and the Sharvin resistance\cite{ref18}. Using a Sc/Sm/Sc
junction with a wide constriction, we obtain $D=0.59$ and $Z=0.83$ for our
SQPC. Plugging $D=0.59$, $\Delta_{0}=\Delta_{\text{Nb}}=1.27$ meV, and
$\tau_{0}=0.32$ ps (obtained with $L_{\text{ch}}=300$ nm and $v_{\text{f}%
}=\frac{\hbar}{m^{\ast}}\sqrt{2\pi n_{s}}=9.52\times10^{5}$ m/s) into Eq.
(\ref{eq02}) gives $I_{\text{c}}$ of 125 nA for $\alpha=1$ and 5.9 nA for
$\alpha=D/4\pi=0.047$. The experimental $I_{\text{sw0}}$ ($=10.3$ nA) lies
between these two values, and it can be explained by assuming a $V_{\text{g}}%
$-independent $\alpha$ with a value of 0.082. However, the absence of any
pronounced peaks in our experimental data implies that no significant
interference takes place and that $\alpha$ should be \symbol{126}1 instead of
small $\alpha$ value. If we use $\alpha=1$, the model gives $I_{\text{c}}%
\sim125$ nA for a single channel, which is one order of magnitude larger than
the experimental $I_{\text{sw0}}$.

The large discrepancy can be resolved by taking into account the proximity
layer at the 2DEG/Sc interface\cite{ref19}, which has a pair potential called
a minigap $\Delta_{\text{mg}}$. To estimate $\Delta_{\text{mg}}$, we fit the
temperature dependence of $I_{\text{sw}}$ using the theoretical model proposed
by Kulik and Omelyanchuk (KO-2)\cite{ref20a, *ref20b}. In the model, the
current-phase relation and critical current are described as $I_{\text{s}%
}(\varphi)=\frac{e\Delta_{0}}{\hbar}\sin\left(  \varphi/2\right)  \tanh\left[
\frac{\Delta_{0}\cos\left(  \varphi/2\right)  }{2k_{\text{B}}T}\right]  $\ and
$I_{\text{c}}=\max\left[  I_{\text{s}}(\varphi)\right]  $, respectively.
Figure \ref{fig04} shows the temperature dependence of the experimental
$I_{\text{sw}}$ along with two calculated $I_{\text{c}}$'s using two different
$\Delta_{0}$ values. Note that $I_{\text{sw}}$ is taken at $V_{\text{g}}=0$ V,
where roughly 15 channels are open, to make sure the Josephson coupling energy
is larger than the thermal energy in the studied temperature range. Looking at
Fig. \ref{fig04}, while the fit using $\Delta_{0}=\Delta_{\text{Nb}}=1.27$ meV
clearly fails to reproduce the experimental data, the best fit is obtained
when $\Delta_{0}=0.154$ meV, which we regard as $\Delta_{\text{mg}}$. Now,
$e\Delta_{\text{mg}}/\hbar$ gives $I_{\text{c}}=38$ nA, still larger than but
of the same order as $I_{\text{sw0}}$. Note that we use $e\Delta_{\text{mg}%
}/\hbar$ instead of Eq. (\ref{eq02}) for calculating $I_{\text{c}}$ because we
do not know $D$, which could be drastically altered from 0.59 once the minigap
is formed. This simplification overestimates $I_{\text{c}}$ and could account
for the remaining difference. Finally, we would like to point out that a
similar $\Delta_{\text{mg}}$ value was obtained in a system similar to our
SQPC from the density of state spectra of the proximity layer formed at the
Nb/In$_{0.8}$Ga$_{0.2}$As interface\cite{Deon}.%

In regard to the $V_{\text{g}}$ dependence of d$I/$d$V$ presented in Fig.
\ref{fig03}(b), a clear stepwise change is also observed and the experimental
data reasonably follow the fitted line. The striking part is that step height
$g_{0}$ is equal to $2.7G_{0}$, which is larger than the quantized conductance
$G_{0}$. This enhancement is a consequence of the multiple ARs in the
ballistic Sc/Sm/Sc junction\cite{ref18}. In contrast to a normal QPC, in which
a limited number of transport modes are available under a finite voltage bias,
the Andreev-reflected quasiparticles in the SQPC carry charges through
fictitious electron or hole bands without exerting additional voltage. An SQPC
is an ideal platform for demonstrating this effect because the conductance is
determined not by the contact resistance at the Sm/Sc interface but by the
number of 1D channels. Closely looking at the data in Fig. \ref{fig03}(b), we
notice that the experimental d$I/$d$V$ deviates from the fitted line and
exhibits a dip structure superimposed on the plateaus. This is most notable
for the first plateau ($n=1$), but other plateaus ($n=4,5,6$) also exhibit the
feature in a more subtle way. The unexpected dip is a characteristic feature
in the low-bias regime. The inset of Fig. \ref{fig03}(b) shows d$I/$%
d$V$-$V_{\text{g}}$ at 2 mV, in which the anomalous dip structure is flattened
and conventional conductance quantization in units of $G_{0}$ is recovered.
Since the charge transport is governed by the AR in the low-bias regime, these
results indicate that the AR probability changes with $V_{\text{g}}$ (or the
position of the Fermi level in the QPC). The origin of this anomalous behavior
is not clear, and further study is necessary to elucidate it.

\section{Conclusion}

A Sc/Sm hybrid SQPC made of an In$_{0.75}$Ga$_{0.25}$As QPC with Nb electrodes
is examined. The quantization of $I_{\text{c}}$ is demonstrated by the
staircase variation of $I_{\text{sw}}$. The staircase variation persists down
to the single-channel regime, providing the first unambiguous demonstration of
a Josephson junction with a single ballistic channel using Sc/Sm hybrid SQPCs.
Although this quantized critical current has already been reported, the
results presented in this paper prove it in a much clearer fashion. Beyond the
experimental proof of quantized $I_{\text{c}}$, the realization of SQPC,
especially single channel operation, opens up interesting possibilities for
the application in quantum information processing. A pair of Andreev levels in
a single channel SQPC forms a doublet state that can be utilized as a quantum
bit\cite{ref28}. Recently, a spectroscopy analysis of the Andreev levels in an
Al-based MCBJ was performed, and its phase-dependent energy levels were
successfully demonstrated\cite{ref29}. The studied SQPC allows us to precisely
control the channel number, which is an advantage in realizing the Andreev
level qubits.

\begin{acknowledgments}
This work was supported by the "Topological Quantum Phenomena" (No. 22103002)
Grant-in Aid for Scientific Research on Innovative Areas from the Ministry of
Education, Culture, Sports, Science and Technology (MEXT) of Japan.
\end{acknowledgments}

\bibliographystyle{apsrev4-1}
\bibliography{acompat,SQPC-irie-2nd}

\begin{thebibliography}{35}
\bibitem [{ref()}]{ref11}%
  \BibitemOpen
  \href@noop {} {}\bibinfo {note} {The actual layer structure is, from the
  bottom to the surface, an undoped In$_{0.52}$Al$_{0.48}$As buffer (200 nm), a
  Si 4$\times$ 10$^{18}$ cm$^{-3}$ doped layer (6 nm), an undoped
  In$_{0.52}$Al$_{0.48}$As spacer (10 nm), a composite 2DEG layer consisting of
  In$_{0.53}$Ga$_{0.47}$As/In$_{0.75}$Ga$_{0.25}$As /In$_{0.53}$Ga$_{0.47}$As
  (2.5/8/5 nm), an undoped In$_{0.52}$Al$_{0.48}$As layer (3 nm), an undoped
  InP surface layer (5 nm).}\BibitemShut {Stop}%
\end{thebibliography}

\end{document}